\documentclass[preprint,5p,times,twocolumn]{elsarticle}

\usepackage{amssymb}
\usepackage{amsmath}
\usepackage{graphicx}
\usepackage{xcolor}
\usepackage{caption}   
\usepackage{subfigure} 
\usepackage{array}
\usepackage{subcaption}
\usepackage{algorithm}
\usepackage{algorithmic}
\usepackage{booktabs}
\usepackage{multirow}
\usepackage{pifont}

\begin{document}

\begin{frontmatter}



\title{Early-MFC: Enhanced Flow Correlation Attacks on Tor via \\Multi-view Triplet Networks with Early Network Traffic}


\author[1]{Yali Yuan}
\ead{yaliyuan@seu.edu.cn}

\author[1]{Qianqi Niu}
\ead{qianqiniu@seu.edu.cn}

\author[2]{Yachao Yuan*}
\ead{chao910904@suda.edu.cn}

\address[1]{Southeast University, School of Cyber Science and Engineering, Nanjing, China}
\address[2]{Soochow University, School of Future Science and Engineering, Suzhou, China}

\cortext[cor1]{Corresponding author}

% \affiliation{organization={},%Department and Organization

\begin{abstract}
Flow correlation attacks is an efficient network attacks, aiming to expose those who use anonymous network services, such as Tor. Conducting such attacks during the early stages of network communication is particularly critical for scenarios demanding rapid decision-making, such as cybercrime detection or financial fraud prevention. Although recent studies have made progress in flow correlation attacks techniques, research specifically addressing flow correlation with early network traffic flow remains limited. Moreover, due to factors such as model complexity, training costs, and real-time requirements, existing technologies cannot be directly applied to flow correlation with early network traffic flow. In this paper, we propose flow correlation attack with early network traffic, named Early-MFC, based on multi-view triplet networks. The proposed approach extracts multi-view traffic features from the payload at the transport layer and the Inter-Packet Delay. It then integrates multi-view flow information, converting the extracted features into shared embeddings. By leveraging techniques such as metric learning and contrastive learning, the method optimizes the embeddings space by ensuring that similar flows are mapped closer together while dissimilar flows are positioned farther apart. Finally, Bayesian decision theory is applied to determine flow correlation, enabling high-accuracy flow correlation with early network traffic flow. Furthermore, we investigate flow correlation attacks under extra-early network traffic flow conditions. To address this challenge, we propose Early-MFC+, which utilizes payload data to construct embedded feature representations, ensuring robust performance even with minimal packet availability. Simulation results demonstrate that Early-MFC reduces packet requirements by 80\% compared to the state-of-the-art DeepCoFFEA system, while Early-MFC+ maintains formidable attack efficacy even when constrained to only the first ten packets of each flow.
\end{abstract}
%arbitral 


% \begin{highlights}
% \end{highlights}

\begin{keyword}



flow correlation \sep
multi-view \sep
triplet network \sep
Tor

\end{keyword}

\end{frontmatter}

\setlength{\parindent}{2em}
\section{Introduction} \label{sec:Introduction}
\begin{figure}[t] 
    \centering % 居中
    \includegraphics[width=0.5\textwidth]{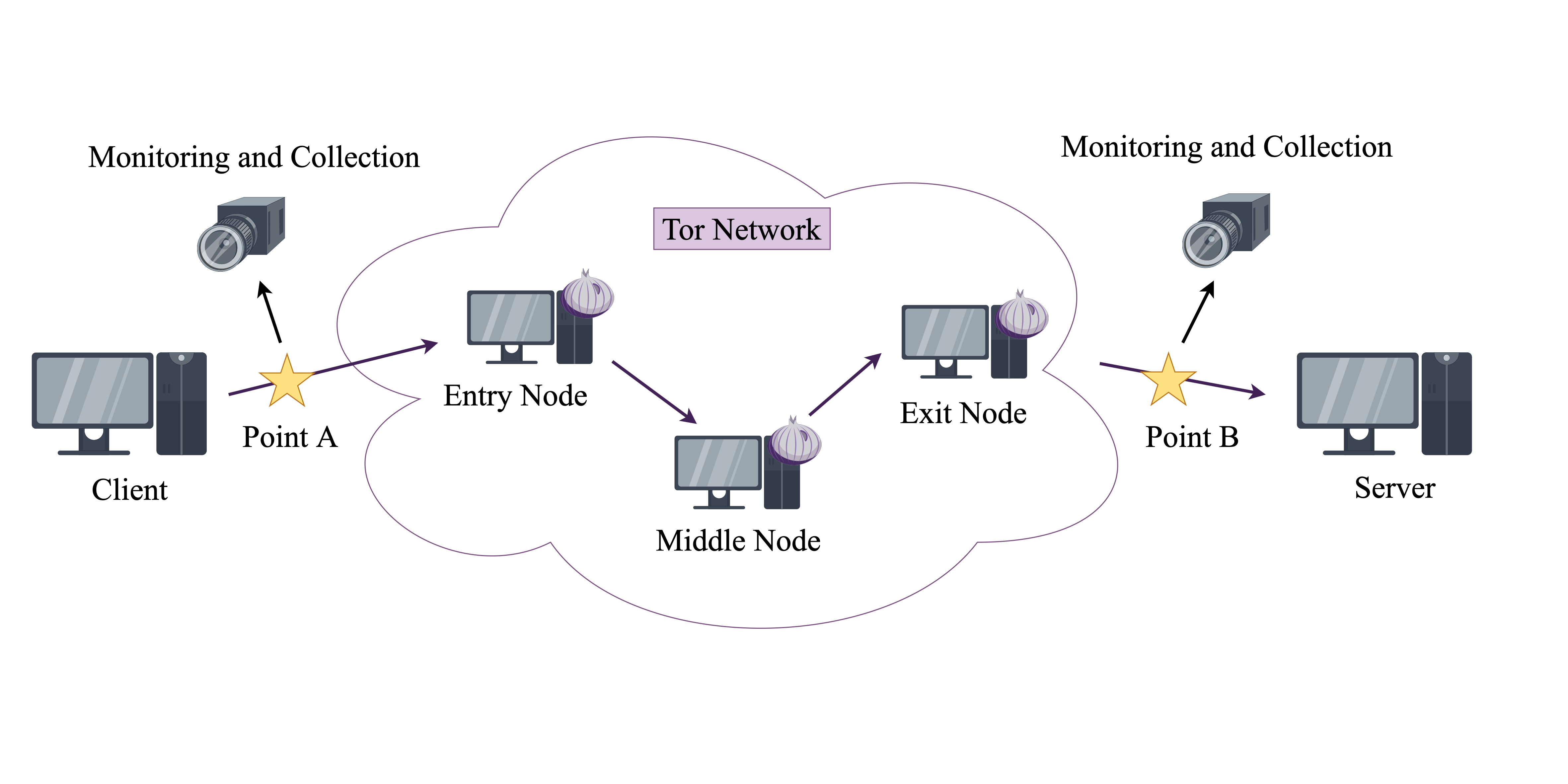}
    \caption{The structure of a Tor network. Data is collected at points A and B in the picture.}
    \label{fig:tor}
\end{figure}

The Onion Router (Tor) \cite{syverson2001towards} is one of the most widely recognized anonymous networks, serving millions of users daily. Tor enables users to access the internet while mitigating the risk of their online activities being monitored or logged by third-party entities, such as Internet Service Providers (ISPs). Tor operates on a distributed network architecture, where internet traffic is routed through a series of volunteer-operated servers, referred to as nodes or relays \cite{mani2018understanding}. When a user initiates a connection through the Tor network, their data is encrypted and routed through a randomly selected sequence of at least three nodes, forming a secure circuit. A schematic representation of this process is illustrated in Figure \ref{fig:tor}. This multi-layered encryption and routing mechanism effectively obscures the user's IP address and browsing activities from potential network observers, such as Internet Service Providers (ISPs) or surveillance agencies.

With the continuous advancement of attacks technology, the deanonymization of the Tor network has become increasingly viable. The development of flow correlation attacks has significantly bolstered this process. Flow correlation attack reveals user identity or monitors user behavior by matching the traffic sent by the client and the traffic received by the server by observing and analyzing network traffic characteristics like sending time and packet size.

In this paper, we investigate flow correlation with early network traffic, which refers to conduct flow correlation as early as possible  during traffic data transmission. This approach is particularly critical in scenarios that demand rapid response and real-time processing. By enabling timely identification and detection of illicit activities within the Tor network, flow correlation with early network traffic facilitates swift intervention and prevention measures. For instance, it allows for the prompt detection of abnormal user behaviors, thereby enabling the implementation of appropriate countermeasures. Overall, flow correlation with early network traffic enhances response times and supports administrators in making timely and informed decisions.

Traditional flow correlation attacks typically employ non-machine learning techniques, such as time-frequency domain analysis \cite{zhu2005flow}, Spearman correlation coefficient \cite{raptor} or compressed traffic analysis \cite{nasr2017compressive}. These methods often necessitate manual feature selection and tuning, a process that heavily depends on large-scale datasets to ensure feature effectiveness and model accuracy. For instance, RAPTOR requires the collection of 100 MB of traffic data within a 5-minute window for each intercepted flow. In recent years, with the development of deep learning, especially the application of convolutional neural network (CNN) in flow correlation, has led to significant progress in this domain. Notable examples include DeepCorr \cite{deepcorr} and DeepCoFFEA \cite{deepcoffea, guan2023flowtracker}.  These models leverage the capability of CNN to automatically learn and extract features from data, enabling them to process complex datasets and identify critical features for classification and recognition tasks, thereby enhancing overall accuracy. Among these works, DeepCoFFEA have further incorporated contrastive learning techniques to compare similarities between different samples. This approach improves the model's ability to identify and distinguish key features, ultimately leading to more robust performance. Nevertheless, despite reductions in the number of required packets, existing methods still necessitate the collection of a substantial volume of traffic data—approximately 1,000 packets per intercepted flow (900 for DeepCorr and around 1,000 for DeepCoFFEA)—to train deep learning models. This extensive data requirement ensures that the models can effectively capture the complex patterns and relationships within the flow data. However, constrained by factors such as model complexity, training costs, and real-time requirements, existing attacks are generally not directly applicable for effective flow correlation in the early stages of data flow generation. This limitation is particularly pronounced in application scenarios that demand rapid response and real-time processing, where effective countermeasures cannot be implemented in a timely manner when swift reactions are required. For example, this includes the timely and accurate detection of cybercrime or the prevention of financial fraud. Therefore, there is a pressing need to develop an enhanced flow correlation attack capable of achieving high performance even with early network traffic.

% ensuring high performance in flow correlation despite the absence of certain data to meet practical application requirements is an urgent challenge that needs to be addressed.
% achieving high performance of flow correlation in the absence of some data to meet the requirements of practical applications is an urgent problem to be solved.
%Therefore, that can maintain high accuracy even when data is severely missing. 
%This is crucial for enhancing the applicability and robustness of models in real-world scenarios.
 
To address these challenges, we propose Early-MFC, a novel framework for flow correlation attack with early network traffic based on multi-view triple networks. The proposed framework aims to extract traffic features from multiple perspectives by integrating multi-view data and leveraging multiple model architectures. It ultimately achieves high-accuracy early orrelation analysis in early stage by fusing the embedding representations from each model. Firstly, the framework first extracts comprehensive traffic features from multiple perspectives to obtain a holistic view of the traffic data. In the process it employs different model architectures to learn and represent these features effectively. Specifically, CNN is utilized to extract spatial features from payload data, while Long Short-Term Memory network (LSTM) capture temporal features from Inter-Packet Delays (IPDs). And then, the framework fuses the embedding representations from these models using techniques such as metric learning and contrastive learning. By comparing the similarities between different samples, the framework ensures that similar flows are mapped closer together in the embedding space, while dissimilar flows are positioned farther apart thereby enhancing the model's discriminative ability and robustness. Through this multi-view and multi-model fusion approach, Early-MFC achieves high-precision traffic correlation in the early stages of flow, providing effective technical support for scenarios that demand rapid response and real-time processing. We validate the performance of Early-MFC through extensive experiments on the latest Tor network dataset. The results demonstrate that Early-MFC achieves higher accuracy, true positive rate (TPR), and lower false positive rate (FPR) compared to state-of-the-art methods, while requiring significantly fewer traffic packets.

In summary, the main contributions of this paper are as follows:
%For the first problem, the training cost are reduced by lessening the number of training samples while ensuring 3MFC's performance. 
%we shorten the training cycle while reducing the number of training samples and ensuring that the model performance is not affected. 
%For the second problem, we leverage the advantages of flow data modalities. The convolutional neural networks and Long Short-Term Memory method are used to extract the critical features (i.e.,  raw features and IPD features) of varied modalities to reveal the potential connections between flows.
%analyze flow from multiple perspectives, combined with convolutional neural networks and Long Short-Term Memory to extract raw flow features and size sequence data features and reveal potential connections between the flow.
%For the third problem, we design a feature reconstruction network to solve the data missing problem. The model's performance is ensured even when facing a data missing rate of up to 90\%, which makes our solution more suitable for practical security applications. 

% In summary, the main contributions of this paper are as follows:
\begin{itemize}
    \item This paper proposes a novel flow correlation attack framework, Early-MFC, designed specifically for early network traffic. Implemented using multi-view triplet networks, Early-MFC extracts features from multiple perspectives of early network traffic flows and maps them into a shared embedding space. This embedding space is then processed by a triplet network, which effectively distinguishes correlated flows from non-correlated flows in early network traffic. By leveraging multi-view feature extraction and triplet network optimization, Early-MFC significantly reduces the training cost of the model while also minimizing the requirements for traffic data collection and storage.
    
    \item To effectively conduct flow correlation attacks with extra-early network traffic, we propose an enhanced feature reconstruction network, named Early-MFC+. This network leverages payload data to construct embedded feature representations, thereby achieving robustness and effectiveness attack performance even when constrained to the initial ten packets of each flow. 
    
    \item This paper compares a variety of attacks and evaluates their performance on different datasets. The experimental results demonstrate that Early-MFC performs extraordinarily well in flow correlation attacks meanwhile Early-MFC reduces the packet requirement by 80\% compared to DeepCoFFEA. The proposed Early-MFC+ demonstrated an accuracy of 93\% on the latest flow correlation dataset, while simultaneously reducing data packet requirements by 99\% in comparison to the DeepCoFFEA.
\end{itemize}

The rest of this paper is organized as follows. Section \ref{sec:RelatedWork} reviews related studies on flow correlation. Section \ref{sec:Framewrok} shows an overview of Early-MFC, Early-MFC+ and presents details of each key technology. Section \ref{sec:ExperimentalSetup} and Section \ref{sec:ExperimentResults} discuss the experimental setup and achieve results. Section \ref{sec:Conclusion} summarizes this paper.

\section{Related Work} 
\label{sec:RelatedWork}
In this section, we first provide an overview of the primary setup and key considerations in a flow correlation scenario. We then delve into a detailed review of the related work on flow correlation within the context of the anonymous network.

\subsection{Threat Model}
Flow correlation attacks are a type of passive analysis method specifically targeting low-latency anonymity networks such as Tor. These attacks aim to link users with their potential destinations by determining whether an entry flow and an exit flow within the network correspond to the same communication connection. To achieve this, the attacks compute the correlation between pairs of flows to determine a relationship between the flow entering and exiting the network. As shown in Figure \ref{fig:tor}, attackers can intercept traffic at points of the network (e.g., on the entry or exit nodes) without full control of the network. Specifically, we monitor the entry flow from a client to the entry node at point $A$, while we monitor the exit flow from the exit node to the server at point $B$. The flow collected at point $A$ is called $\textit{entry}_i$, and at point $B$ is called $\textit{exit}_i$. Previous studies on flow correlation attacks \cite{deepcorr, deepcoffea} concentrate on a correlation metric $M(\textit{entry}_i, \textit{exit}_i)$, and computing $M(\textit{entry}_i, \textit{exit}_i)$ for every pair of observed traces within a specific time frame. The attacker concludes that the flows are correlated if $M(\textit{entry}_i, \textit{exit}_i)$ is greater than or equal to a threshold \(\tau\).

In anonymous networks like Tor, user anonymity is safeguarded primarily through encryption and traffic mixing. However, flow correlation attacks can circumvent these protective measures by analyzing traffic characteristics, thereby compromising user anonymity. As a result, flow correlation attacks pose a significant threat to the integrity of anonymity in networks such as Tor\cite{dingledine2004tor}. 

\subsection{Flow Correlation Attacks}
Zhu et al. \cite{zhu2005flow} posed two classes of correlation approaches that were considered, namely time-domain and frequency-domain methods. Based on their threat model and known strategies in existing mix networks, they conducted experiments to analyze the performance of mixes. Murdoch et al. \cite{murdoch2007sampled} assumed that the traffic entering and leaving a Tor network passed through attacker-controlled Internet Exchange Points and attempted to match target traffic entering the network with traffic leaving the network. Song et al. \cite{song2013anonymize} developed a de-anonymize attack method based on traffic analysis and chose the {time, size} as features for the k-means algorithm to mine the correlation between the first hop traffic and last hop traffic of Tor. RAPTOR \cite{raptor} discussed the anonymity of Tor networks and how it can be threatened by attackers who can de-anonymize Tor clients by observing traffic on both ends of the Tor communication channel. Nasr et al. \cite{nasr2017compressive} proposed the core idea of compressive traffic analysis, which is to compress traffic features, and perform traffic analysis operations on such compressed features instead of on raw traffic features. Compressive traffic analysis leveraged linear projection algorithms from compressed sensing, an active area within signal processing, to compress traffic features. DeepCorr \cite{deepcorr} leveraged advanced deep learning models to learn flow correlation functions of complex networks like Tor. Cheng et al. \cite{cheng2018flow} proposed an enhanced random forest-based DDoS attack detection method based on genetic algorithm optimization of flow correlation degree features. Guan et al. \cite{guan2019empirical} accessed a self-built server through an SSH proxy and Tor successively, capturing data flows generated by different pluggable transports and upper applications. Then, identification and correlation techniques based on various machine learning algorithms are used to break anonymity. Guan et al. \cite{guan2020restor} designed a pre-processing model called ResTor to perform the noise reduction before actually correlating entering and exiting flows. ResTor treats the byte accumulation sequences smoothed at fixed intervals as fitting targets, and takes advantage of the stacked auto-encoder architecture to remove noise in two phases. Li et al. \cite{li2021attcorr} proposed AttCorr that takes raw traffic features of flow pairs as inputs, including packet size, flow direction, and inter-packet delay, and used a multi-head attention mechanism to capture the flow information and noise in Tor. DeepCoFFEA \cite{deepcoffea} used deep learning to train a pair of feature embedding networks. It uses amplification, dividing traffic into short windows and using votes from these windows to reduce false positives. Guan et al. \cite{guan2023flowtracker} proposed an improved flow correlation attacking model named FlowTracker, which used stacked autoencoders to learn mapping relationships between relevant input and output cumulative representations and filter out network noise.

\begin{figure*}[t] 
    \centering % 居中
    \includegraphics[width=1\textwidth]{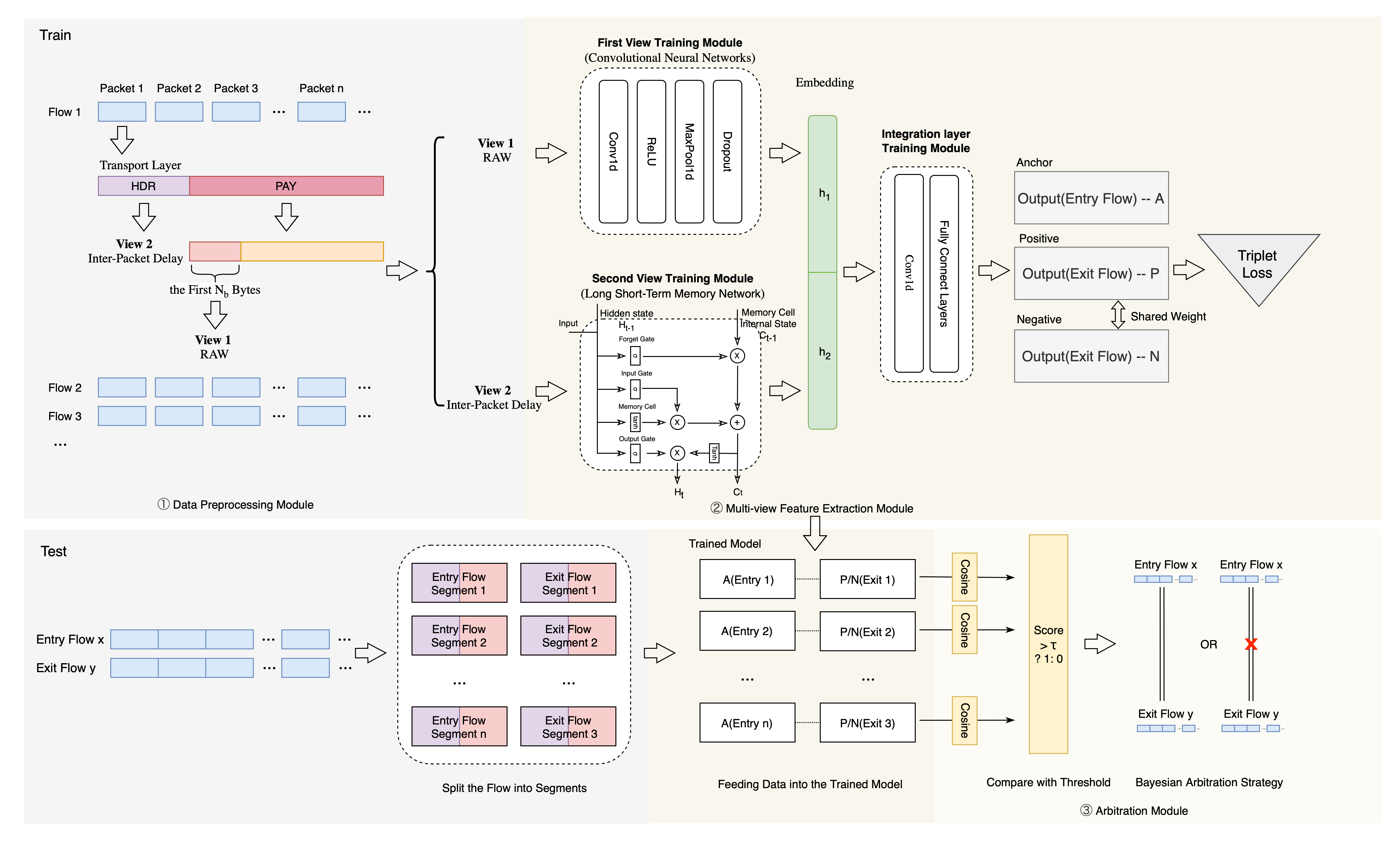}
    \caption{The figure describes the framework of Early-MFC, which is roughly divided into three blocks. Block \ding{192} describes the stage in which we process data, dividing the pacp data into available multi-view input data, and then putting the multi-view data into module \ding{193} for comparative learning triplet training. Block \ding{194} makes the final arbitration through the Bayesian rule.}
    \label{fig:frame}
\end{figure*}

\section{Multi-view Deep Learning–based Flow Correlation Framework} \label{sec:Framewrok}
In this section, we present the proposed Early-MFC and Early-MFC+ frameworks, designed for flow correlation in scenarios involving early and extra-early network traffic. We begin by providing an overview of the Early-MFC and Early-MFC+ framework. And then we delve into a detailed description of each component, elucidating their specific roles and functionalities within the overall framework. 

\subsection{Framework Overview}
As illustrated in Figure \ref{fig:frame}, the proposed Early-MFC framework contains three components, i.e., data preprocessing module (see \ding{192} in Figure \ref{fig:frame}), multi-view feature extraction module, (see \ding{193} in Figure \ref{fig:frame}), and final arbitration module (see \ding{194} in Figure \ref{fig:frame}).

Specifically, within the data preprocessing module, we concentrate on segmenting network packet files (i.e., pcap traces) through slicing operations, thereby decomposing the network traffic into packet segments. Subsequently, these segmented data are transformed into multi-view representations. By processing data from multiple perspectives, we are able to capture and understand the characteristics of network traffic in a more comprehensive manner. In the multi-view feature extraction module, the pre-processed multi-view data is input into a meticulously designed model architecture. Within this architecture, each view's data is utilized to optimize its corresponding sub-model. These sub-models operate independently on their assigned data, generating individual outputs. Subsequently, these outputs are passed to an integration layer, which is tasked with fusing the predictions from different views. This fusion enables Early-MFC to fully leverage data from diverse sources and perspectives, thereby enhancing its capability to learn network flow patterns even with limited data. The triplet loss function is employed for model optimization. In the arbitration module, following the approach inspired by DeepCoFFEA \cite{deepcoffea}, a data flow is divided into k time slots.For each pair of time slots, k correlation metrics are computed and compared against a predefined threshold. The final decision is then made using a Bayesian method based on these comparisons.

\begin{figure}[t] 
    \centering % 居中
    \includegraphics[height=10cm]{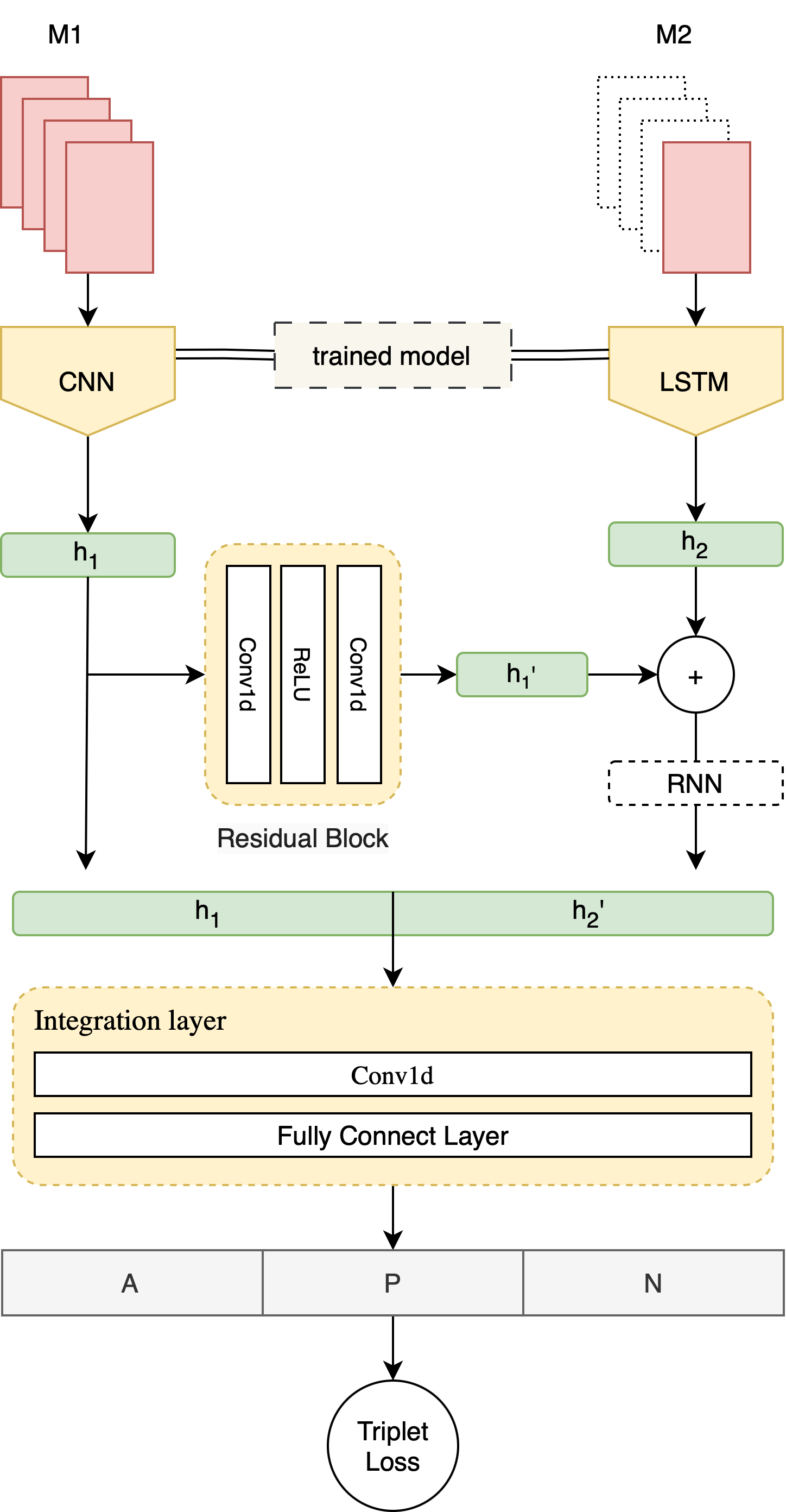}
    \caption{The structure of the feature reconstruction network.}
    \label{fig:missing}
\end{figure}

Moreover,we propose a new feature reconstruction network, as illustrated in Figure \ref{fig:missing}, which replaces Block \ding{193} in Figure 2.
This approach addresses the challenges of conducting flow correlation at the extra-early stage of network traffic generation, which is a requirement in real-world applications such as the detection of cybercrime or the prevention of financial fraud. By leveraging the available payload data to replenish and reconstruct the Inter-Packet Delays (IPDs) feature, the proposed method effectively maintains high accuracy even when using extremely limited traffic information.

\subsection{Data Preprocessing Module}
Following \cite{aceto2021distiller, aceto2019mimetic}, Early-MFC eemploys a multi-view feature extraction strategy that simultaneously captures flow characteristics from packet payload content and inter-packet delays (IPDs), ensuring high correlation accuracy even under early network traffic constraints.  Through this data processing approach, the Early-MFC is capable of extracting the characteristics of flows from multiple perspectives, which enables it to effectively capture both local information carriers and global statistical features in traffic patterns. In the subsequent section, we provide a detailed exposition on the methodology for extracting the required data from pcap files.

Firstly, the pcap file is processed to generate input data, which is categorized into two distinct types \cite{aceto2021distiller}: (a) the initial segment of the transport-layer payload (PAY), consisting of the first $N_b$ bytes; and (b) the informative fields of the protocol headers (HDR) from the first $N_p$ packets. The first category is depicted in a Decimal format from raw hexadecimal, organized byte by byte. The (b) encompasses (i) the byte count of a transport-layer payload, (ii) TCP window size, which is zeroed for UDP packets, (iii) the time intervals between packet arrivals - IPDs, and (iv) the packet direction, indicate by a binary value of either 0 or 1, all pertaining to the initial $N_p$ packets. In this paper, we choose IPDs as the second data view, the specific reasons are described in detail in Section \ref{sec:ExperimentResults}. The following formula expresses the relationship between flows and packets:
$$ F = \{p_1, p_2, ..., p_n | p = (\textit{HDR},\textit{PAY}), n = 200\},$$
where $F$ is the flow and $p$ is the packet.We call (a) as RAW and (b) as IPDs. 

RAW data contains the original payload of the network flow, preserving the most complete information that serves as the key local features in traffic pattern analysis. As these data have not undergone any preprocessing, they retain all potential features that may be utilized to identify the traffic type. This unprocessed raw payload can reflect the essential characteristics of network traffic to the greatest extent. We extract the first 10 packets from each flow and utilize the first 80 bytes of each packet as the RAW input. By employing the payload as the input data for the model, we are able to capture the inherent structural features of the data packets and the potential patterns embedded within the raw byte sequence.

Through IPDs, attackers can analyze the temporal relationships and statistical properties between consecutive packets, thereby revealing the overarching behavioral patterns of network flows. Through the analysis of IPDs, one can deduce the macroscopic characteristics of flows. In this paper, we extract the top 200 IPDs from each flow for analysis. The temporal intervals between packets provide insights into the global characteristics of the traffic, while the payload information reveals the local characteristics. The information derived from these two complementary perspectives can collectively enhance the accuracy of the model.

To ensure uniformity in input dimensions, data instances exceeding the specified size are truncated, while undersized instances are zero-padded to match the predefined byte size $N_b$ or packet size $N_p$. This input normalization process is crucial to mitigate input bias and ensure consistent feature extraction across varying data scales. Truncation encourages the model to learn generalizable feature representations that are robust to variations in data size, while zero-padding preserves structural integrity without introducing artifacts. Together, these techniques ensure that the model's performance remains unbiased and scalable, regardless of differences in input data dimensions.

\subsection{Multi-view Feature Extraction Module}
The module comprises three training components: the first view (RAW) training module, the second view (IPDs) training module, and the integration layer training module.

\textbf{The first view (RAW) training module.} CNN have demonstrated exceptional feature extraction capabilities in the analysis of network traffic data. In this module, high-level feature representations are extracted through stacked convolution and pooling operations. Previous studies \cite{deepcorr, deepcoffea} have shown that CNN performs well in extracting features to analyze Tor flow.
Additionally, existing studies have shown that 1D convolutional layers are often considered to be suitable for processing one-dimensional sequence data, such as time series, audio signals, or one-dimensional text data, while 2D CNN is often considered to be more powerful for processing two-dimensional data like image data. We experimentally evaluate the developed model. Experimental results show that the module based on 1D convolutional layers performs better in generating feature embeddings. In addition, we find that the 1D convolutional layers can more effectively reduce the Triplet loss, while the 2D convolutional layers perform poorly in reducing the Triplet loss. In summary, we use 1D convolutional layers to train data of the first view (RAW data). The block is a sequence model composed of multiple convolutional layers, activation functions (ELU and ReLU), pooling layers (MaxPool1d), and Dropout layers. 
$$ h_1 = \textit{CNN}_{blocks}(M1), $$ 
where h\(_1\) is the feature of M1.

The channel parameters of the convolutional layers of each block gradually increase, indicating that the number of channels of the input and output feature map increases layer by layer. The kernel size and stride of the pooling layer are fixed, but the padding parameter is dynamically calculated by a custom function. Then, a flattened layer and a linear layer (Linear) are used to map the vector to a high-dimensional space. Finally, a fully connected layer is used to fix the output dimension.

\textbf{The second view (IPDs) training module.} LSTM has demonstrated remarkable success in diverse time series-related tasks, such as time series forecasting, speech recognition, natural language processing, and network traffic analysis, underscoring its robust capability in modeling temporal dependencies. IPDs are essentially time series data, reflecting the time interval pattern of packet arrival. LSTM can effectively capture both long-term and short-term dependencies within these intervals, which is a reasonable and effective choice for network traffic analysis.

\begin{algorithm}[t]
\caption{Training Strategy}  
\label{alg:Training}  
\hspace*{0.02in}{\bf Input:}
\\(a) \textit{input\(_a\)}: entry data (b) \textit{input\(_p\)}: exit data (c) \textit{input\(_n\)}: exit data \\Each input contains:
\\(1) multi view 1 (M1): consisting of the first N\(_b\) bytes of PAY \\ (2) multi view 2 (M2): inter-packet delay\\
\hspace*{0.02in}{\bf Output:} 
\\\textit{out\(_a\)}, \textit{out\(_p\)}, and \textit{out\(_n\)} correspond to the output results of input\(_a\), input\(_p\), and input\(_n\), respectively.

\begin{algorithmic}[1] 
\FOR {$i \leftarrow 0$ to \textit{epoch}}          %FOR循环结构
\STATE {convolution operation: $h_1=\mathcal{CNN}(M1)$}
\STATE {regularization technique: $h_1=\mathcal{D}(h_0)$}
\STATE {Long Short-Term Memory: $h_2=\mathcal{LSTM}(M2)$}
\STATE {feature concatenation: $h = h_1+h_2$}
\STATE {integration: $out_a,out_p,out_n\leftarrow \mathcal{F}(\mathcal{C}(h))$}
\STATE {triplet loss:\\
 $Loss = max(0, d(out_a, out_p) - d(out_a, out_n) + \text{margin}) $\\
where margin is the minimum distance margin that we want to maintain.
}
\ENDFOR
\end{algorithmic}  
\end{algorithm}

Initially, an embedding layer is constructed to transform the IPDs into fixed-dimensional vectors. In this model, the IPDs that can be processed is an integer. Following this, LSTM is define, which takes the embedded IPDs vector as input. The input dimension is consistent with the output dimension of the embedding layer in the previous row. LSTM is set to be bidirectional and can process forward and reverse sequence data at the same time. The Dropout layer is set between the LSTM layers to avoid model overfitting by adjusting the dropout rate. Using the embedding layer to reshape it into a three-dimensional tensor, and ultimately pass the reshaped input data to the LSTM layer, and get the output of LSTM and the final hidden state and cell state. 
$$ h_2 = \mathit{LSTM(M2)}, $$
where h\(_2\) is the feature of M2. Finally, pass it to a fully connected layer. 

\textbf{Integration layer training module.} The module comprises a convolutional layer and a fully connected layer. Within this framework, feature fusion is conducted on the data output from the first-view data (RAW) training module and the data output from the second-view data (IPDs) training module. The fused features are input into a convolutional layer and subsequently processed through a fully connected layer to produce the final output
$$ output = FC(Conv(Concatenation(h_1,h_2))),$$
where FC is the fully connected layer, and Conv is the convolutional layer. 

The integration of these two modalities empowers the Early-MFC model to characterize network flows from multiple perspectives, thereby enhancing the accuracy and robustness of classification. Specifically, the RAW view concentrates on the granular content of data packets, whereas the IPDs view emphasizes the macroscopic structural features of packet transmission. Through the fusion of features derived from these distinct modalities, the Early-MFC model achieves a more comprehensive understanding of network flow characteristics and establishes a more precise correlation, even in the condition of early network traffic. The overall algorithm is presented in Algorithm \ref{alg:Training}.

\subsection{Triplet Network Loss}
 Within this framework, we employ Triplet Loss as the primary loss function to optimize the model's capacity for embedding representations of input data. Triplet loss works by constructing a triplet (anchor, positive, negative), where the anchor is a reference sample, the positive is a sample similar to the anchor, and the negative is a sample dissimilar to the anchor. The goal of this loss function is to make the distance between the anchor and the positive much smaller than the distance between the anchor and the negative, so as to effectively distinguish samples of different categories in the feature space. The calculation formula for triplet loss is:
$$ L = \sum_{i=1}^{N} \max\left(0, d(a_i, p_i) - d(a_i, n_i) + \text{margin}\right), $$
where d(*,*) represents the distance between samples, a\(_i\) is the anchor point, p\(_i\) is the positive sample, n\(_i\) is the negative sample, and margin is the minimum distance margin that we want to maintain. 

In our approach, triplet loss is utilized to instruct the model to map similar samples to proximate locations within the feature space, while simultaneously driving dissimilar samples to more distant locations. This strategy helps improve performance of the model when dealing with tasks that require distinguishing subtle differences. Ultimately, through adjustment, our framework can effectively use triplet loss to train a model with strong discrimination ability, enabling it to show excellent performance in experiments.

\subsection{Arbitration Module}
\label{}
To mitigate the FPR, DeepCoFFEA incorporates an amplification strategy within a randomization framework. Specifically, the collected network flow is segmented into multiple smaller, overlapping sub-windows, each of which is evaluated independently. The final flow correlation attack results are then derived through a voting arbitration mechanism among the windows. Voting arbitration require windows to arbitrate the correct result only when the majority of arbitration are correct. When the relative correctness axiom does not hold, in order to solve this problem, an innovative Bayesian estimation cosine similarity based
arbitration method is proposed, which can select the optimal output result from a probability perspective. Bayesian total probability formula:
$$ P(D) = \sum_{i=1}^{N} P(D|C_i) \times P(C_i), $$
where P(D$|$C\(_i\)) is the likelihood of observing data C\(_i\) under category D. P(C\(_i\))is the prior probability of class C\(_i\). P(D) is the total probability of the data D.

Assume that the cosine similarity information of all windows is W\(_k\), and k represents the result of the kth window. F\(_n\) is the cosine similarity probability distribution for given the flow pair D\(_n\). The flow correlation output result set is y\(_n\)=\{y\(_1\), y\(_2\), …, y\(_n\)\}. We define a threshold $\tau$, if the maximum value in F\(_n\) exceeds this threshold $\tau$, then y is marked as correlation. At the same time, the threshold can be dynamically updated according to historical arbitration. The cosine similarity of the feature vectors of flows under attack after adjusting by the Bayesian method is expressed as follows: 
$$ y = argmax\,P(F_i|D_n), $$

Here, the Bayesian algorithm is adopted to select the arbitration result, that is, to select the category with the maximum probability value as the final arbitration.

\subsection{Extra-Early Flow Correlation}
\label{}
To address the critical demand for real-time flow correlation of extra-early network traffic in specialized environments, we propose a novel feature reconstruction network framework that effectively caters to this requirement. We establish an exceedingly ambitious objective of achieving high-precision flow correlation by utilizing merely 10\% of the packets required by the Early-MFC model in each flow. To achieve this objective, we developed the Early-MFC+ network, which leverages payload data to compensate for the incomplete feature representation of IPDs. This enhancement improves the model's adaptability to extra-early flow correlation tasks. This approach significantly reduces the amount of data required for model training, offering a more practical solution for flow correlation tasks in specialized environments. The implementation details of Early-MFC+ are as follows:

First, we define two modalities: let payload data as M1, while IPDs is M2. We perform independent view feature extraction. We input M1 and M2 into two independent training models in 3.3, which can extract features from their respective data.
\begin{equation}
h_1=\text{CNN}(M1), h_2=\text{LSTM}(M2)),
\end{equation}
where $h_1$ represents the features extracted from M1, and h\(_2\) represents the features extracted from M2. Next, we use a residual block, which is designed to enhance feature representation and improve the model's learning ability for complex data patterns by introducing residual connections. We input h\(_1\) into for further processing to obtain $h_1'$:
\begin{equation}
 h_1' = h_1 + \text{ReLU}(W_2 * \text{ReLU}(W_1 * h_1 + b_1) + b_2).
\end{equation}

After being processed by the residual block, the obtained $h_1'$ is embedded into $h_2$ as $h_2'$, which is the reconstructed embedding of the IPDs. Before that, in order to better merge the features $h_1'$ and $h_2$, the merged $h_1'$ and $h_2$ are input into the RNN for further feature extraction and refinement, and finally, $h_2'$ is obtained:
\begin{equation}
h_2' =\text{RNN}(h_1'+h_2).
\end{equation}

The process of concatenating $h_1$ and $h_2'$ and outputting them to an integration layer, which consists of a 1D convolution and a fully connected layer, can be represented by the following steps and formulas:
\begin{equation}
O = \text{FC}(\text{ReLU}(\text{Conv1D}([h_1;h_2' ],K,F,P,S)),W_fc,b_fc),
\end{equation}
where K is the number of convolution kernels, F is the size of each convolution kernel, P is the padding size, S is the stride, and W\(_fc\) and b\(_fc\) are the weights and biases of the fully connected layer, respectively.

In our experiments, we simulated scenarios with varying packet counts and demonstrated how our feature reconstruction network effectively maintains model accuracy. Specifically, when Early-MFC+ uses only 10\% of the packets required by Early-MFC, our model can still achieve 93\% accuracy in flow correlation by leveraging payload information.

\section{Experimental Setup}\label{sec:ExperimentalSetup}
\subsection{Dataset}
Publicly available datasets exist for flow correlation research, such as DeepCorr \cite{deepcorr} and DeepCoFFEA \cite{deepcoffea}. However, these datasets do not adequately account for the multi-view characteristics of the data. To address this limitation, we collect data following the data collection protocols outlined by the DeepCorr \cite{deepcorr} dataset. Specifically, we employ Tor version 0.4.8.11 for our experiments. We deploy clients across multiple isolated virtual machines to generate and capture Tor flow. We randomly select over 6,000 websites from the Alexa Top 20,000 list as targets for data collection. During the browsing process, we utilize 990 distinct Tor circuits. Consistent with the approach taken in DeepCorr, we employ various guard nodes when constructing these circuits, relying primarily on public Tor relays to ensure realism in flow patterns. The entry flow is captured using tcpdump at the client, while we make our exit Tor flow tunnel through our own SOCKS proxy server and utilize tcpdump on the proxy server to capture Tor exit flow.

We employ the obfs4 pluggable transport to obfuscate flow. obfs4 encrypts and transforms the flow exchanged between the client and the guard node, making it less recognizable to adversaries \cite{deepcoffea}. obfs4 provides Inter-Arrival Time (IAT) randomization, which improves flow indistinguishability by introducing random delays between packets and perturbing their timing intervals. To emulate realistic network user behavior, we alternate between two obfs4 modes during data collection: IAT mode "on", which introduces randomized delays during data transmission, and IAT mode "off", which preserves the original flow characteristics without obfuscation \cite{deepcorr}. 

The data collection process is divided into two phases: an initial two-week period, followed by an additional one-month collection conducted after a one-month interval. This methodology allow us to analyze the temporal impact on training. Each sample is captured over a 30-second interval, although less data than the full 30-second capture is typically utilized during practical analysis. Each website is visited three times, resulting in a total of 39,426 data pairs. Our dataset is publicly available, with 80\% of the captured data designated as training data and the remaining 20\% reserved for evaluating the model's performance. Finally, we filter out flows with packet counts less than 8, which is shorter than DeepCorr and DeepCoFFEA. DeepCorr mainly filtered 300, 400, and 500 packets, while DeepCoFFEA filtered 70.
Meanwhile, we also consider the legal and ethical aspects of the dataset. During the data collection process, we ensur that the Tor network is not overloaded and no illegal content is browsed during the experiment.

\subsection{Hyperparameters}
Hyperparameter selection is a critical aspect of model training, as it directly influences the model's learning capability and ultimate prediction accuracy. To optimize hyperparameters effectively, we employ a random search strategy, which explores the parameter space through random sampling. This approach enable us to evaluate a broad range of hyperparameter combinations within a relatively short timeframe. Additionally, cross-validation techniques are incorporated to ensure that the selected hyperparameter configurations not only achiev optimal performance on the training set but also maintain robustness on unseen data.

Our hyperparameter optimization process target several key components of the model framework and training strategy. First, we fine-tuned the learning rate, which determines the step size for parameter updates during each epoch as the model converges toward the minimum of the loss function. Selecting an appropriate learning rate is vital: an excessively high value risks overshooting the optimal point, while a value that is too small can result in slow convergence.

The choice of optimizer is another critical consideration. We experiment with several optimizers, including Adam, Stochastic Gradient Descent (SGD), and RMSprop, to assess their effects on convergence speed and stability. Each optimizer has unique parameters, such as momentum in SGD and the beta coefficients in Adam, which are carefully adjust to enhance performance. Furthermore, we analyze the effect of the number of training epochs to identify an appropriate balance that prevents both underfitting and overfitting. Model-specific hyperparameters, such as the number of layers and neurons per layer for CNN and LSTM components, are also included in the tuning process. By evaluating various configurations, we assess their impact on the model's ability to capture intricate patterns within the data.

The results of our hyperparameter tuning process are summarized in Table \ref{tab:hyperparameter}, which provides the optimal values for key hyperparameters such as learning rate, batch size, optimizer choice, number of epochs, and architecture-specific parameters. These selections enable the model to achieve its best performance.

\renewcommand{\arraystretch}{1.2}
\begin{table}[t]
    \centering
    \caption{Choice of hyperparameter.}
    \begin{tabular}{p{0.25\textwidth}p{0.1\textwidth}}
        \hline
        Hyperparameters & Value \\  \hline
        Extra-Early Packets Number & 200 \\ 
        Extra-Early+ Packets Number & 20/100/180 \\
        batch size & 64 \\ 
        output size & 32 \\ 
        epoch number & 30 \\ 
        learning rate & 0.001 \\ 
        optimizers & Adam \\ 
        number of window & 5 \\ \hline
    \end{tabular}
\label{tab:hyperparameter}
\end{table}

\subsection{Comparison of attack methods}
This paper compares several current mainstream attack methods. This study selects four attack methods as comparison objects. We evaluate them from multiple dimensions. Then, through quantitative and qualitative analysis, we show their performance in different datasets. (1) DeepCorr \cite{deepcorr}, a deep learning-based attack model that uses CNN to learn and identify correlation patterns in Tor network flow. (2) DeepCoFFEA \cite{deepcoffea}, combines metric learning and amplification techniques to improve the accuracy of attacks and reduce false positives. (3) Compressive Analysis \cite{nasr2017compressive}, this method uses cosine distance as a statistical distance metric to perform correlation analysis on the compressed packet time series of flow. (4) Raptor \cite{raptor}, Raptor uses the Spearman correlation coefficient as a correlation indicator, focuses on analyzing the IPDs to identify correlations between flow, and is a statistical-based method.

\renewcommand{\arraystretch}{1.2}
\begin{table*}[!t]
    \centering
    \caption{Comparison results with advanced models.}
    \begin{tabular}{c c c c c c c c c c c}
        \hline
            Negative Sample Number & &  9 &  &  & 99 &  & & 190 &  &  \\  
           \cmidrule(lr){1-2}\cmidrule(lr){3-5}\cmidrule(lr){6-8}\cmidrule(lr){9-11}
        Dataset & Method & ACC & TPR & FPR & ACC & TPR & FPR& ACC & TPR & FPR\\  \hline
          & DeepCorr & 90.05 & 82.29 & 0.21& 87.95 & 79.16 & 0.32& 83.76 & 69.79 & 0.21\\  
          & DeepCoFFEA & 97.38 & 95.83 & 0.04 & 96.85 & 94.79 & 0.08 & 92.30 & 93.75 & 0.07\\  
        D1 & Raptor & 60.87 & 59.94 & 2.94 & 56.96 & 57.75 & 7.05 & 55.49 & 40.62 & 8.64\\
          & CTA & 69.63 & 69.58 & 4.32 & 63.05 &  62.46 & 4.78 & 58.77 & 62.97 & 6.71\\  
          & Early-MFC & 98.42 & 97.91 & 0.01 &98.42 & 98.95 & 0.02 & 96.85 & 97.91 & 0.04   \\  \hline
          & DeepCorr &86.91 & 76.04 & 0.31& 86.38 & 74.54 & 0.35& 79.58 & 59.37 & 0.52 \\  
          & DeepCoFFEA  & 96.33 & 94.79 & 0.11 & 94.76 & 90.26 & 0.23  & 91.25 & 92.70 & 0.53\\  
        D2 & Raptor & 58.43 & 56.25 & 3.93 & 57.70 & 55.31 & 4.67 & 53.40 & 71.87 & 6.52\\
          & CTA & 78.01 & 63.54 & 3.86 & 73.54 & 61.66 & 4.10 & 57.05 & 58.91& 6.58 \\  
          & Early-MFC & 97.40 & 96.87 & 0.01 & 96.85 & 97.91 & 0.03& 96.40 & 93.57 & 0.06\\  \hline
          & DeepCorr & 85.86 & 75.00 & 0.47& 76.96 & 64.58 & 1.05 & 75.39 & 52.08 &1.38\\  
         & DeepCoFFEA & 93.71 & 89.58 & 0.21 & 91.25 & 89.58 & 0.67 & 91.09 & 84.37 & 0.79  \\  
         D3 & Raptor & 53.07 & 53.21 & 13.26 & 52.81 & 52.60 & 15.02 & 52.28 & 52.33 & 15.92\\  
          & CTA & 64.86 & 62.10 & 6.63 & 60.01 & 62.65 & 7.17 & 56.25 & 52.82 & 7.92\\ 
          & Early-MFC & 97.90 & 98.95 & 0.04 & 96.33 & 94.79 & 0.10& 96.85 & 95.83 & 0.16\\  \hline
    \end{tabular}
    \label{tab:Comparison}
\end{table*}

\subsection{Evaluation metrics}
The True Positive Rate (TPR), also refers to as Sensitivity or Recall, and the False Positive Rate (FPR) are widely used metrics in flow correlation and deep learning to evaluate the performance of correlation models. TPR quantifies the proportion of actual positive samples correctly identified by the model, while FPR measures the proportion of negative samples incorrectly classified as positive. These metrics are fundamental for assessing a model's ability to distinguish between positive and negative classes.

The formula for calculating the True Positive Rate is:
$$TPR= \frac{TP}{(TP+ FN)},$$
and the formula for calculating the False Positive Rate is:
$$FPR= \frac{FP}{(FP+ TN)},$$
where: 
\begin{itemize}
    \item TP (True Positives): The count of positive samples correctly predicted by the model.
    \item FN (False Negatives): The count of positive samples incorrectly predicted as negative.
    \item FP (False Positives): The count of negative samples incorrectly predicted as positive.
    \item TN (True Negatives): The count of negative samples correctly predicted by the model.
\end{itemize}

Achieving a high TPR often comes with the risk of an increased FPR, as the model may incorrectly classify more negative samples as positive. Conversely, minimizing the FPR may require raising the correlation threshold, potentially lowering the TPR by misclassifying some true positive samples as negative. Thus, when evaluating a model, it is essential to strike an appropriate balance between the TPR and FPR to achieve optimal performance in real-world scenarios.

Accuracy is another commonly used metric for evaluating correlation models. It represents the proportion of correctly classified samples—whether positive or negative—relative to the total number of samples, providing an overall assessment of model performance. Accuracy is define as:
$$Acc= \frac{TP+ TN}{TP+ TN+ FP+ FN}.$$

In this study, we use a combination of these metrics—TPR, FPR, and accuracy—to comprehensively evaluate and compare our model's performance against existing models presented in related works. This multi-faceted approach ensures a thorough assessment of the model's capabilities in flow correlation tasks.

\section{Experiment Results} \label{sec:ExperimentResults}
In this section, we present and analyze experimental results that compare the performance of the Extra-Early with contemporary state-of-the-art attack techniques. In addition, we will also conduct ablation experiments and sensitivity analysis to further explore the role of each component of the model and its robustness to input changes.
\subsection{Experiment settings}
In this study, we implement our deep learning models using the PyTorch framework, leveraging CUDA technology to optimize computational efficiency. Model training and evaluation are conducted on an NVIDIA V100 GPU with 32GB of video memory, ensuring sufficient resources for handling large-scale datasets and complex model architectures. To minimize bias in our analysis, we employ a random sampling strategy to create distinct data subsets. From the total dataset of 39,426 samples, we randomly select 10,382 samples to form Dataset D1. Subsequently, 25,670 samples are randomly drawn from the remaining data to constitute Dataset D2. Finally, 37,680 samples are selected to create Dataset D3. To enhance the model's generalization capability, the dataset is collected at different times and utilized distinct Tor circuits, enabling the model to adapt to unknown data in varying conditions.

To validate the robustness and practical applicability of our model, we intentionally preserve the original state of the dataset, refraining from data cleaning or preprocessing. This decision aim to maintain the natural distribution of the data, allowing the model to encounter noise and outliers. By doing so, we could assess the model's ability to operate effectively in real-world environments where data imperfections are common. This approach also facilitate an evaluation of the model's resilience to noise and its capability to generalize from unprocessed, authentic data.

To mitigate potential bias and ensure adaptability to diverse data distributions, training and testing samples are selected through random sampling. This randomized selection process help prevent overfitting to specific data patterns, enabling a more comprehensive assessment of the model's performance.

Through these evaluations, we aim to demonstrate that our model is not only theoretically effective but also practically reliable and robust in real-world applications. This dual focus on theoretical soundness and empirical feasibility underscores the potential of our approach for deployment in complex and dynamic operational environments.

\subsection{Comparison to the State-of-the-Art}
In this section, we present a comprehensive comparative analysis to highlight the differences between our approach and existing methods. For method that provide public code,we directly use those code to reproduce their results in my experimental environment, ensuring that all comparisons are conducted under the same conditions. For methods without publicly available code, we implement the corresponding models based on the methodological descriptions provided in the respective papers. During this process, we carefully adhere to the algorithms and procedures outlined in the original publications to ensure fidelity.

To promote transparency and reproducibility in scientific research, we have decided to make our code and datasets publicly accessible. This initiative allows other researchers to validate our findings and leverage our dataset for their own investigations.

The experimental results, summarize in Table \ref{tab:Comparison}, underscore the superiority of our method in terms of both ACC and TPR. Specifically, our model continues to exhibit substantial performance advantages even under conditions where the number of data packets is limited. Experimental results show that the Early-MFC method shows significant advantages in multiple datasets (D1, D2, D3) and different numbers of negative samples (9, 99, 190). In terms of accuracy, Early-MFC reaches 98.42\% on the D1 dataset, and stabilizes at 97.40\% and 97.90\% on the D2 and D3 datasets, respectively, which is significantly better than other methods such as DeepCoFFEA. At the same time, the TPR of Early-MFC is close to 99\%, and the FPR is always maintained at an extremely low level (0.01\% to 0.16\%), which is much lower than methods such as Raptor and CTA. In addition, as the number of negative samples increases, the performance of Early-MFC decreases the least, showing its strong robustness to data imbalance problems. These results fully demonstrate the efficiency and practicality of Early-MFC in flow correlation analysis tasks.

We attribute these high levels of performance to several factors. The superior performance of Early-MFC is mainly due to its multi-view feature fusion strategy, which can simultaneously capture the local features of the payload content and the global features of the temporal dynamics, thereby maintaining high discrimination ability in early network traffic. In addition, Early-MFC further enhances the robustness of feature representation through the joint optimization of contrastive learning and metric learning. This enables it to effectively distinguish the behavioral patterns of different flows even in environments with a high proportion of negative samples.

The results confirm that Early-MFC achieves superior performance across diverse conditions, demonstrating significantly higher TPR for any given FPR compared to other methods. Especially in the early traffic analysis scenario, Early-MFC can achieve high-precision flow correlation analysis with only a small number of packets, providing strong support for real-time network security protection.

The experimental results demonstrate that Early-MFC achieves outstanding performance across all datasets (D1, D2, D3), significantly outperforming existing methods. Particularly in early-stage traffic analysis scenarios, Early-MFC is capable of achieving high-precision flow correlation analysis with only a limited number of packets, providing robust technical support for real-time network security protection.

\renewcommand{\arraystretch}{1.2}
\begin{table}[t]
    \centering
    \caption{Result of Early-MFC+ effectiveness.}
    \begin{tabular}{c c c c c }
        \hline
         Method & Packets Number  &  ACC & TPR & FPR \\  \hline
          & 180 & 97.38 & 95.83 & 0.010\\  
         Early-MFC+&100 & 96.33 & 94.79 & 0.021\\  
          & 20 & 93.71 & 89.58 & 0.021\\  \hline
    \end{tabular}
    \label{tab:missingtab}
\end{table}

\renewcommand{\arraystretch}{1.2}
\begin{table}[t]
    \centering
    \caption{Result of ablation experiment}
    \begin{tabular}{p{0.14\textwidth}p{0.08\textwidth}p{0.08\textwidth}p{0.08\textwidth}}
        \hline
        & ACC & TPR & FPR \\  \hline
         RAW + IPDs & 98.42 & 97.91 & 0.010 \\ 
        RAW  & 91.09 & 84.37 & 0.021\\  
        IPDs  & 70.15 & 53.12 & 1.263 \\  \hline
    \end{tabular}
    \label{tab:ablation}
\end{table}

\begin{figure*}[ht] 
    \centering % 居中
    \includegraphics[width=\textwidth]{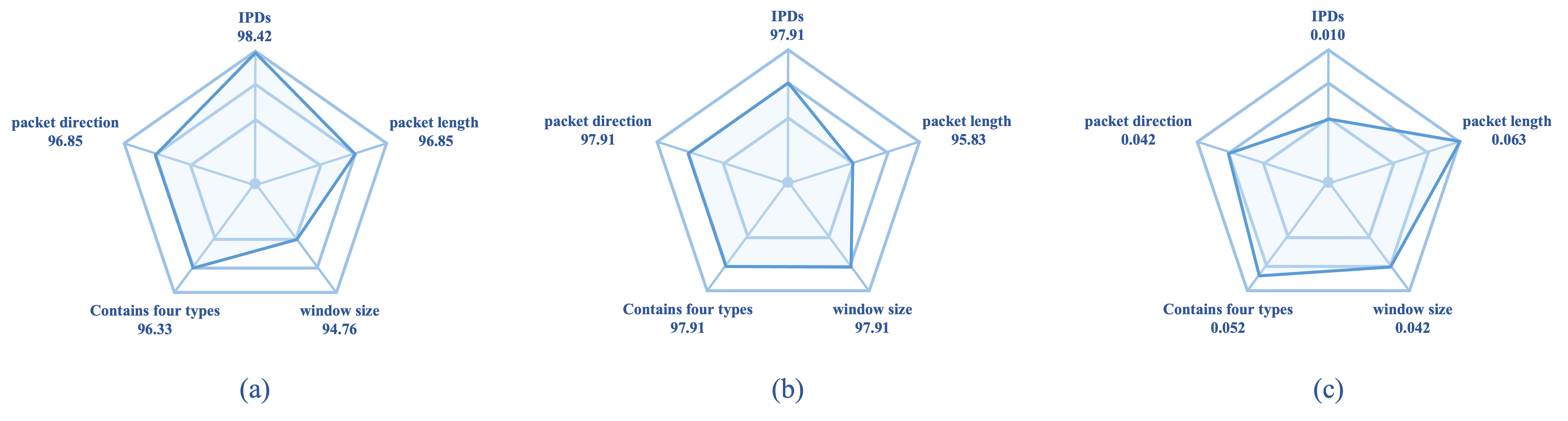}
    \caption{The figure describes the results of different choice of the second view under the same experimental conditions. (a) is the result of ACC, (b) is the result of TPR, and (c) is the result of FPR.}
    \label{fig:choose}
\end{figure*}

\subsection{Performance of Early-MFC+}
In this experiments, we simulated scenarios with different numbers of packets. As the capturing traffic time decreases, the number of packets decreases accordingly. We conducted experiments on the D3 dataset with scenarios involving 10\%, 50\%, and 90\% of the packets required by Early-MFC, aiming to replicate realistic flow correlation with extra-early network traffic and assess the model's robustness and adaptability under these conditions. The results are summarized in Table \ref{tab:missingtab}.

The results reveal that when 180 packets (90\%*Early-MFC) are used, the performance of the Early-MFC+ model is only 1\% lower than that of Early-MFC. With 100 packets (50\%*Early-MFC), the flow correlation performance decreases slightly, by approximately 1\% compared to the 180-packet condition. This demonstrates the robustness of the Early-MFC+ model in extra-early network traffic condition.

When only 20 packets (10\%*Early-MFC) are available, the performance of Early-MFC+ decreases significantly, dropping around 3\% compared to the 100-packet condition. Despite this reduction, the model maintains an accuracy of over 90\%. This notable performance indicates that even with extra-early traffic, the model can still effectively integrate and utilize information from the remaining payload, enhancing overall prediction accuracy.

These experimental findings validate the effectiveness of the proposed network and highlight the potential of multi-view learning in addressing challenges of flow correlation extra-early network traffic. The demonstrated robustness and adaptability of the model are particularly valuable for practical applications in network security, ensuring reliable flow analysis in extra-early network traffic condition.

\subsection{Ablation Analysis and Sensitivity Analysis}
In this section, we perform sensitivity analysis and ablation analysis of the Early-MFC by varying key parameters and observing their impact on attack effectiveness.

\textbf{Ablation analysis.} We conduct ablation experiments on the D3 dataset to evaluate the importance of multi-view within the model. To assess the specific contribution of different views to the model’s performance and verify the effectiveness of multi-view fusion, we compare the performance of three configurations: (a) The fully complete views. (b) A model trained using only the first view (RAW). (c) A model trained using only the second view (IPDs).
The experimental results, summarize in Table \ref{tab:ablation}, show that the full-view model achieves an accuracy of 98.42\% on the test set, significantly outperforming the single-view models. Specifically, the accuracy of the model using only the first view is 91.09\%, while the model using only the second view achieves an accuracy of 70.15\%. 

From a theoretical perspective, in early network traffic, single-view data offers only limited perspectives and information. Consequently, the model may struggle to capture the complete structure and semantics of the data, rendering it susceptible to noise, overfitting, and biases in data distribution. In contrast, multi-view models are capable of more effectively capturing the intrinsic structure of the data, thereby demonstrating stronger generalization ability and mitigating the impact of noise inherent in single-perspective data.

The experimental results demonstrate that, even with limited packets, the application of multi-view fusion can significantly preserve high accuracy in flow correlation by leveraging the complementary information from different views.

\begin{figure}[t] 
    \centering % 居中
    \includegraphics[width=0.9\linewidth]{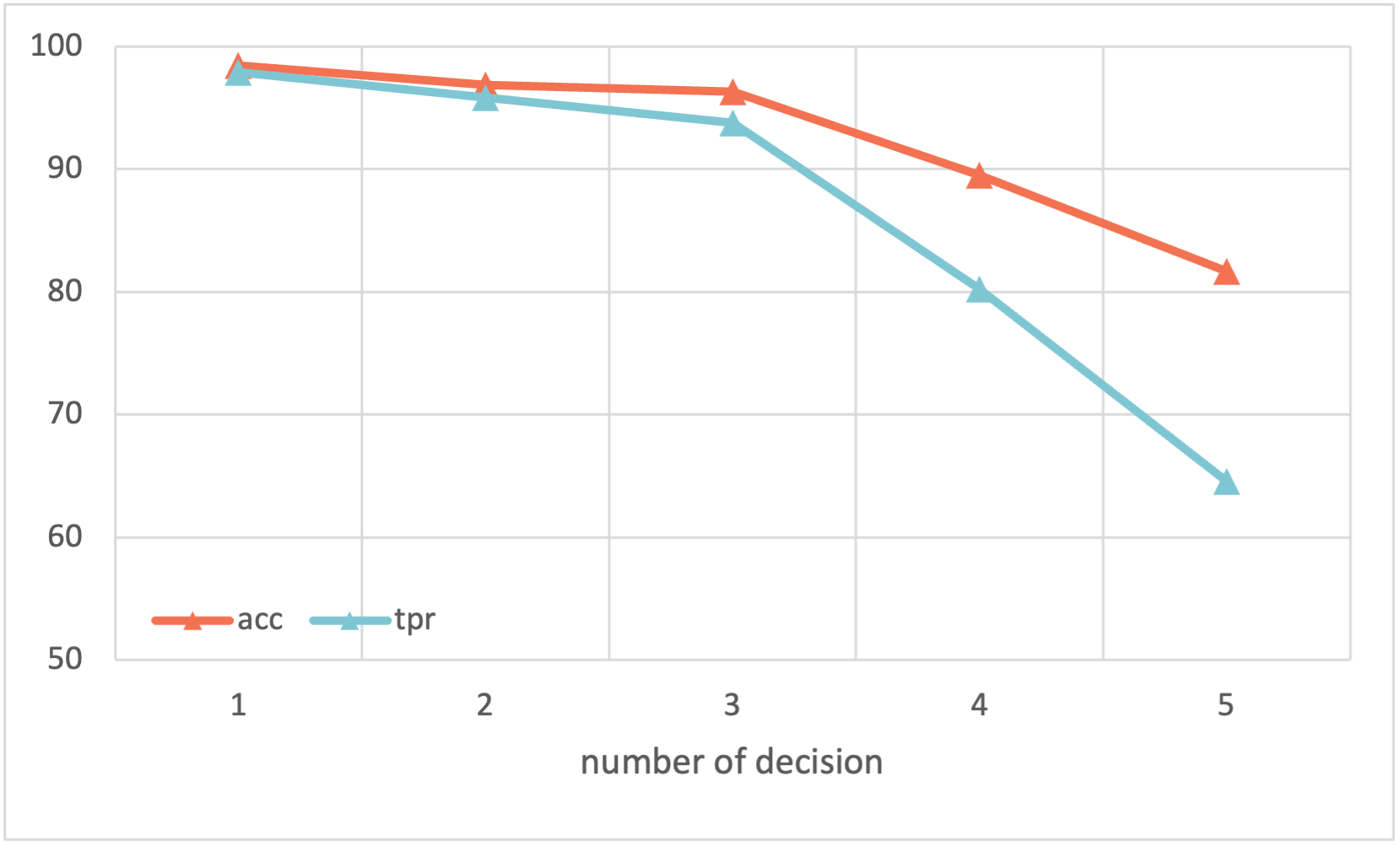}
    \caption{The figure describes the results of different arbitration}
    \label{fig:decision}
\end{figure}

\textbf{Sensitivity analysis -- the choice of the second view.} We analyze existing works in the field of multi-view traffic classification, introducing the common feature sets typically chosen as the second view as follow. These works often utilize a four-tuple feature set, which includes: (i) the number of bytes of the transport layer payload, (ii) the TCP window size, (iii) the time interval between packet arrivals-- IPDs, and (iv) the direction of the packet, represented by binary values 0 or 1. We conduct independent experiments using each of these features to evaluate their individual contributions and effectiveness for flow correlation tasks. The results of these experiments are illustrated in Figure \ref{fig:choose}.

The experimental results show that when we combine the LSTM network with (iii) IPDs, the accuracy at test time exceeds that of (i) the number of bytes in the transport layer payload, (ii) the TCP window size, and (iv) the direction of the packet. In comparison with these features, IPDs combination with the LSTM network exhibits a more pronounced synergistic effect. We posit that this enhanced performance is likely attributable to the distinctive internal memory mechanism of the LSTM network, which is particularly adept at processing and forecasting time-series data.The gating mechanism in its design (including input gate, forget gate, and output gate) enables it to learn the temporal correlation of data and effectively capture long-term dependencies. Consequently, in the context of network traffic analysis, the LSTM network can leverage the temporal information features within the IPDs feature to discern the evolving patterns of network traffic over time.

While the traditional four-tuple feature set has historically been considered foundational for multi-view network traffic classification due to its widespread application, our research shows that the IPDs feature alone achieves slightly higher accuracy—approximately 1.6\% better than the traditional four-tuple set. When using only the IPDs feature, the model achieves comparable performance in terms of TPR relative to the four-tuple feature set, indicating that both feature sets are similarly effective at identifying positive classes. However, in terms of FPR, the IPDs feature outperforms the four-tuple feature set, demonstrating a lower rate of mis-correlation for negative samples.

This finding suggests that the IPDs feature contains information valuable for traffic analysis, which traditional four-tuple features may not fully exploit. The IPDs feature can serve as a more robust and informative view, particularly when combined with time-series processing models like LSTM. This finding underscores the potential of focusing on specialized features like IPDs for improved performance in network traffic analysis tasks.

\textbf{Sensitivity analysis -- arbitration number.} In our experiments, we use a Bayesian mechanism to enhance the arbitration-making ability of the model. We combine multiple prediction results to improve the overall classification performance. We change the number of models participating in the arbitration, that is, set different arbitration thresholds of 1, 2, 3, 4, and 5 to test the specific impact on the model performance.

Our experimental results, as shown in Figure \ref{fig:decision}, show that in the model with a Bayesian mechanism, the best performance is achieved when the arbitration threshold is set to 1. This finding is consistent with the conclusions drawn from the Bayesian method mentioned in Section 3 of the paper.

\section{Conclusion} \label{sec:Conclusion}
In summary, this paper introduce Early-MFC, an innovative flow correlation attack framework based on multi-view triplet networks, which effectively addresses the need for flow correlation within early network traffic. By leveraging a multi-view approach that processes the payload and IPDs of network flows using CNN and LSTM networks respectively, our model effectively extracts features from a limited number of packets, thereby preserving high accuracy. Through comparative experiments with the current state-of-the-art models, Early-MFC performs extraordinarily well in flow correlation attacks meanwhile Early-MFC reduces the packet requirement by 80\% compared to DeepCoFFEA. Results show that the multi-view data fusion strategy has significant effectiveness in flow correlation.We also introduce Early-MFC+, a novel framework designed for flow correlation with extra-early network traffic, and evaluate its performance on the latest flow correlation datasets. The experimental results demonstrate that, despite requiring 99\% fewer packets than DeepCoFFEA, Early-MFC+ achieves an accuracy rate of 93\%, highlighting its potential utility for network security. Looking ahead, this work pioneers a novel direction in the research of early network traffic analysis and deposit a significant theoretical and technical foundation. In future endeavors, we may further investigate the application of early network traffic analysis in domains such as encrypted traffic analysis, anomaly detection, and network intrusion prevention.

\section{Acknowledgements} 
This work was supported in part by Natural Science Foundation of Jiangsu Province (Grant No. SBK2023041256) and in part by the National Natural Science Foundation of China (Grant No. 62302097).



\bibliographystyle{elsarticle-num} 
\bibliography{ref}


%\begin{thebibliography}{00}



% \end{thebibliography}
\end{document}